# Anoxygenic photosynthesis and the delayed oxygenation of Earth's atmosphere

Kazumi Ozaki[1,2,3†], Katharine J. Thompson[4], Rachel L. Simister[4], Sean A. Crowe[1,4], and Christopher T. Reinhard[1,2*]

[1]School of Earth and Atmospheric Sciences, Georgia Institute of Technology, Atlanta, GA 30332
[2]NASA Astrobiology Institute, Mountain View, CA 94043
[3]NASA Postdoctoral Program, Universities Space Research Association, Columbia, MD 21046
[4]Depts of Microbiology & Immunology and Earth, Ocean, & Atmospheric Sciences, University of British Columbia, Vancouver, BC V6T 1Z3
[†]Current address: Department of Environmental Science, Toho University, Funabashi, Chiba 274-8510, Japan

*To whom correspondence should be addressed. E-mail: chris.reinhard@eas.gatech.edu

**The emergence of oxygenic photosynthesis created a new niche with dramatic potential to transform energy flow through Earth's biosphere. However, more primitive forms of photosynthesis that fix $CO_2$ into biomass using electrons from reduced species like Fe(II) and $H_2$ instead of water would have competed with Earth's early oxygenic biosphere for essential nutrients. Here, we combine experimental microbiology, genomic analyses, and Earth system modeling to demonstrate that competition for light and nutrients in the surface ocean between oxygenic phototrophs and Fe(II)-oxidizing, anoxygenic photosynthesizers (photoferrotrophs) translates into diminished global photosynthetic $O_2$ release when the ocean interior is Fe(II)-rich. These results provide a simple ecophysiological mechanism for inhibiting atmospheric oxygenation during Earth's early history. We also find a novel positive feedback within the coupled C-P-O-Fe cycles that can lead to runaway planetary oxygenation as rising atmospheric $pO_2$ sweeps the deep ocean of the ferrous iron substrate for photoferrotrophy.**

## Introduction

The large-scale oxygenation of Earth's atmosphere ~2.3 billion years ago (Ga) signaled one of the most profound biogeochemical shifts in Earth's history[1,2], and occurred as a direct result of photosynthetic oxygen production. It is widely assumed that the ubiquity of water as an electron donor for oxygenic photosynthesis would have conferred an enormous competitive advantage to the first oxygenic phototrophs, facilitating their rapid domination of the photosynthetic niche following their emergence[3,4]. Paradoxically, evidence for oxygenic photosynthesis can be found as much as a billion years before the first large-scale oxygenation of Earth's atmosphere[5-7], and atmospheric $O_2$ has remained well below the present atmospheric level (PAL) for as much as ~90% of Earth's history[1,8,9]. Evidence from Earth's rock record thus suggests a protracted interval of low-oxygen atmospheric chemistry following the evolution of the oxygen-evolving complex in the first oxygenic phototrophs.

Prior to the emergence of oxygenic photosynthesis, the photosynthetic niche would have been populated exclusively by anoxygenic phototrophs[10,11]. Today, anoxygenic phototrophs proliferate in sunlit anoxic environments where they utilize a wide-range of electron donors including dihydrogen, hydrogen sulphide, thiosulphate, elemental sulphur and ferrous iron [Fe(II)][12]. Ferrous iron would have been the most widely available electron donor for anoxygenic photosynthesis throughout much of Earth's early history[13-22], but the birth of the oxygen evolving complex in the ancestors of extant cyanobacteria would have created a new photosynthetic niche decoupled from the supply of Fe(II) to illuminated surface ocean waters. This would have initiated fierce competition for light and bioessential elements—most importantly phosphorus (P)[23-27].

In modern aquatic environments, phototrophs compete for available light energy, introduced at the sea surface and attenuated with depth, and dissolved nutrients, largely mixed upward via advection and turbulent diffusion[28]. As a result, organisms that are adapted to low light intensities can compete more effectively for nutrients introduced from deeper in the water column. In the modern ocean and many lakes, for example, low-light-adapted phytoplankton inhabit deep chlorophyll maxima where they exploit nutrient supplies from below, make large contributions to primary production, and strongly diminish nutrient fluxes to the upper photic zone[29].

Anoxygenic phototrophs are especially well adapted to grow at low light levels, using specialized pigments to effectively harness light in spectral regions not commonly available to oxygenic phototrophs[30,31]. Extant anoxygenic phototrophs—including facultatively anoxygenic cyanobacteria, Chlorobi, Chloroflexi, and purple bacteria—commonly position themselves below oxygenic cyanobacteria and algae where they gain access to inorganic electron donors like reduced sulphur species and ferrous iron [Fe(II)][11,32,33]. Residing deeper in the photic zone, modern anoxygenic phototrophs are known to support appreciable primary production[34,35], while weakening the upward fluxes of reduced species and the supply of nutrients to overlying oxygenic cyanobacteria and algae. In particular, members of the Chlorobi are almost ubiquitously better adapted to low light levels than extant oxygenic cyanobacteria[36,37]. Indeed, they photosynthesize at the lowest light levels recorded[37], and have the metabolic capacity to grow at low light through photoferrotrophy[35].

Collectively, these observations based on the physiology, ecology, and modes of extant photosynthesis imply that early anoxygenic phototrophs, growing through photoferrotrophy in Fe-rich (ferruginous) oceans, may have out-competed primitive oxygenic phototrophs for upwelling nutrients[23]. As a first attempt to quantify the effects of this ecophysiological competition on the Earth system, we combine laboratory physiological experiments, genomics, a local reactive transport framework, and global biogeochemical cycle modeling. We find that competition for light and nutrients between different photosynthetic metabolisms represents a critically important feature of Earth's global $O_2$ cycle, modulating biospheric $O_2$ release and atmospheric oxygen when the ocean interior is Fe(II)-rich. Model sensitivity analysis indicates that the basic dynamics of this competition are robust to a wide range of values for key metabolic parameters. These results, combined with observations from the rock record and genomic constraints on the antiquity of phosphorus metabolism, suggest that this control should have been operative for the vast majority of Earth's history.

**Results**
**P uptake potential and its antiquity in a model photoferrotroph.** The first step in our approach involves a series of experiments to map the metabolic potential of the model pelagic photoferrotroph *Chlorobium phaeoferrooxidans* strain KB01, isolated from Kabuno Bay, East Africa[35]. We find that during incubation experiments in the presence of dissolved Fe(II) and phosphate ($PO_4^{3-}$), KB01 grows rapidly and draws down media $PO_4^{3-}$ while photosynthetically oxidizing Fe(II) (Fig. 1), clearly showing robust growth at very low (nM) concentrations of inorganic P. Notably, we find that KB01 continues to grow and oxidize Fe(II) after dissolved $PO_4^{3-}$ concentrations have decreased to asymptotic values below 4 nM, suggesting that KB01 can access Fe-bound $PO_4^{3-}$ for growth (Fig. 1b).

We also find metabolic potential for the uptake of $PO_4^{3-}$ at low concentrations in KB01's genome sequence. KB01 possesses genes that code for the high-affinity $PO_4^{3-}$ uptake system (*Pst*-phosphate specific transport) belonging to the *pho* regulon, which is induced by phosphorus scarcity and is used in extant marine cyanobacteria to mitigate P starvation[38-40]. In KB01, the *pho* regulon is distributed across multiple contigs with redundancy at multiple locations within the genome (Fig. 2). The genome of KB01 also comprises genes that code for an array of enzymes used in P assimilation, breakdown, and recycling under P starved conditions, including alkaline phosphatases and polyphosphate kinases[41,42] (Supplementary Table 1). Combined with the relatively low light requirements for photosynthesis in *Chlorobium* spp., the metabolic potential for P uptake and growth that KB01 displays under low-P conditions suggests that pelagic photoferrotrophs, as represented by KB01, should be effective competitors for available P in an Fe-rich water column setting with mixed modes of photosynthetic growth.

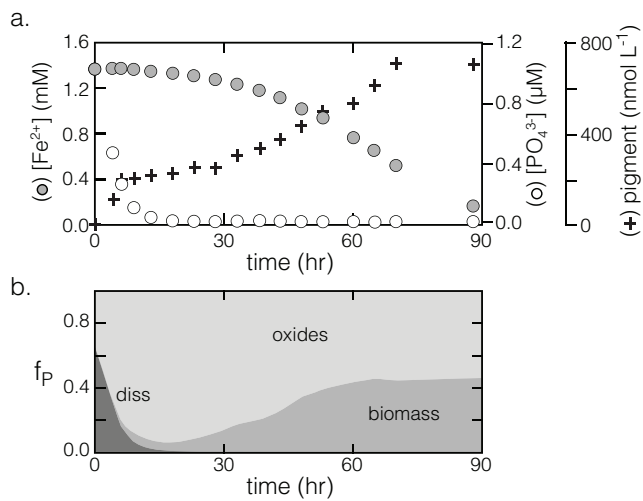

**Figure 1. Physiological data for *Chlorobium phaeoferrooxidans* str. KB01**. (**a**) Time-series data for dissolved phosphate (open circles), dissolved ferrous iron (filled circles), and pigment concentrations (crosses) in batch culture. (**b**) Time-series data for the partitioning of phosphorus between dissolved $PO_4^{3-}$ (diss), Fe-oxide-bound phosphorus (oxides), and phosphorus in photosynthetic biomass (biomass).

As a direct test for the antiquity of high-affinity $PO_4^{3-}$ metabolism in the Chlorobiales, we compare phylogenies of components of the *pho* regulon against highly conserved 16S rRNA gene phylogenies (Fig. 3). A phylogenetic analysis of concatenated PstABC protein sequences, encoding three subunits of the trans-membrane $PO_4^{3-}$ permease, reveals nearly identical branching patterns to those of the 16 rRNA gene, which implies vertical inheritance of the high-affinity $PO_4^{3-}$-uptake system in the Chlorobiales. By extension, this also suggests that the last common ancestor of crown group Chlorobiales likely had similar capacity for $PO_4^{3-}$ metabolism to extant members like strain KB01. Phylogenies of PhoU, which regulates intracellular $PO_4^{3-}$ metabolism, are also broadly congruent with the 16 rRNA gene phylogeny, providing further support for conserved, vertically inherited $PO_4^{3-}$ metabolism in the Chlorobiales.

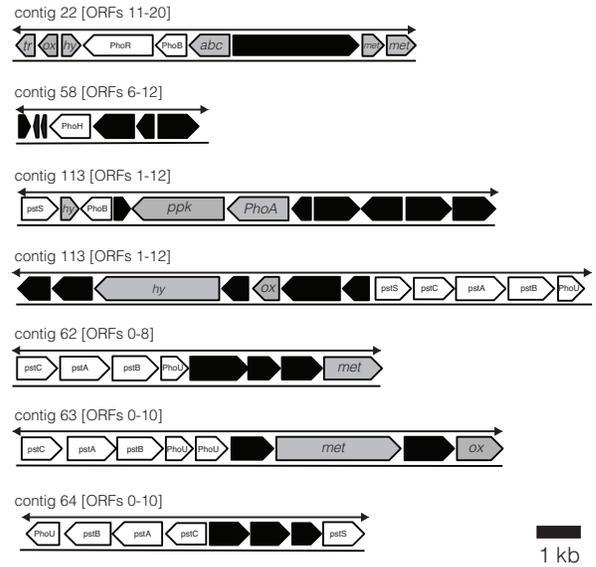

**Figure 2. Phosphorus gene positions in the genome of *Chlorobium phaeoferrooxidans* strain KB01.** Highlighted are the positions of genes involved in high-affinity inorganic phosphorus ($P_i$) uptake (white), including components of the *pho* regulon, alkaline phosphatase (*PhoA*), and polyphosphate kinase (*ppk*). Also shown are positions of transcriptional regulator and transposase proteins (*tr*), membrane, envelope, and transporter proteins (*met*), hydrolases (*hy*), oxidoreductases (*ox*), ABC transporter proteins (*abc*), and other proteins (black) in each open reading frame (ORF). kb = kilobase.

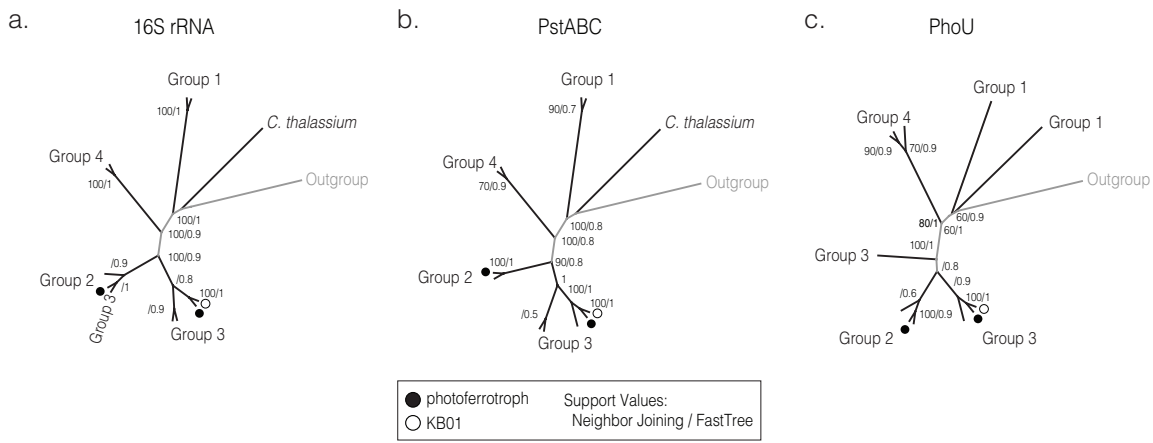

**Figure 3. Comparative molecular phylogeny of 16S rRNA and major genes regulating P metabolism.** Approximately Maximum-Likelihood Trees (FastTree) of concatenated 16S rRNA gene (**a**), PstABC (**b**), and PhoU (**c**). Nodes are annotated with Neighbour Joining (1000x bootstrapped) and FastTree support values. These phylogenies are equivalent to previously established and accepted phylogenies for the Chlorobiales based on the 16S and FMO genes. Maximum Likelihood and Maximum Parsimony yielded trees with similar topology (trees not shown). Clade nomenclature follows recommended taxonomy: Group 1 = *Prostheochloris*, Group 2 = *Chlorobium* 2, Group 3 = *Chlorobium* 3, Group 4 = *Chlorobaculum*. The outgroup is *Rhodothermus marinus* DSM 4252. Circles at the branch tips represent photoferrotrophic organisms and the open circle denotes *Chlorobium phaeoferrooxidans* strain KB01. Note that *C. thalassium* lacks a PhoU gene.

**1-D ecosystem-biogeochemical model.** To explore the potential impact of pelagic photoferrotrophs on nutrient cycling within the photic zone, we embed idealized oxygenic phototrophic and photoferrotrophic organisms in a 1-D advection-diffusion-reaction model meant to represent a generalized photic zone in a ferruginous ocean. Based on well-established models from environmental microbiology and our experimental work with KB01, we specify a simple form of competition between photoferrotrophs and oxygenic phototrophs in which photoferrotrophic growth is parameterized as a function of ambient light, dissolved Fe(II), and nutrient P, while oxygenic phototrophic growth is constrained by both light and dissolved nutrients (see Methods). Light availability is assumed to vary with depth according to a simple exponential (*e*-folding) decay with a specified attenuation length scale. Iron can also be oxidized by dissolved $O_2$, the kinetics of which are specified as a function of ambient $[O_2]$, $[Fe^{2+}]$, pH, temperature, and salinity according to Ref [43]. Scavenging and co-precipitation of inorganic $PO_4^{3-}$ onto/into Fe-oxide mineral phases occurs according to a simple distribution coefficient ($K_d^{FeP}$). Default values for our photic ecology model are given in Supplementary Table 2.

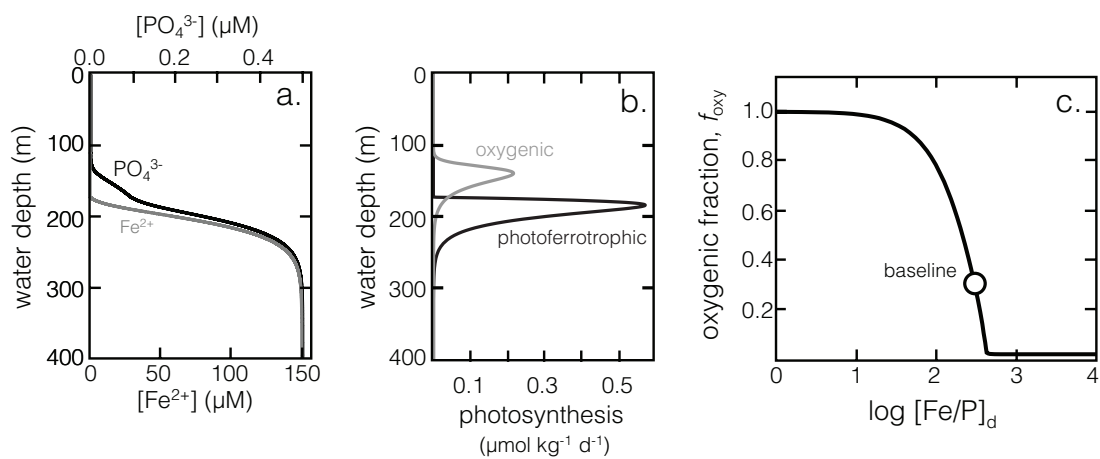

**Figure 4. Representative results from our 1-D competitive photosynthesis model.** (**a**) Water column $PO_4^{3-}$ and $Fe^{2+}$ concentrations; and (**b**) rates of oxygenic and anoxygenic (photoferrotrophic) photosynthesis for our baseline simulation. (**c**) Relative contribution of oxygenic photosynthesis to total water column photosynthesis ($f_{oxy}$) as a function of deep water dissolved Fe/P ratio. The open circle denotes our baseline simulation. Note the log scale in (**c**).

Representative results of our 1-D competitive photosynthesis model are shown in Fig. 4. Because pelagic photoferrotrophs can compete effectively for P at low light levels (i.e., more deeply in the water column), they can strongly attenuate the flux of P to the oxygenic component of the photosynthetic community (Fig. 4a). This results in an emergent tiering pattern in which photoferrotrophs outcompete oxygenic phototrophs for available P until dissolved Fe(II) is exhausted, after which P can break through to shallower oxygenic phototrophs (Fig. 4b). A particularly salient result of our model is the importance of the ratio of Fe(II) to $PO_4^{3-}$ in deepwater ($[Fe/P]_d$) in controlling the relative importance of oxygenic phototrophs to overall system productivity ($f_{oxy}$; Fig. 4c). Above a critical $[Fe/P]_d$ value, pelagic photoferrotrophs effectively eliminate oxygenic phototrophs from the ecosystem by outcompeting them for nutrient P deep within the water column. In contrast, at lower $[Fe/P]_d$ photoferrotrophs become progressively Fe(II)-limited, and oxygenic phototrophs garner a larger fraction of the photosynthetic niche.

Our principal results are very insensitive to the metabolic parameters of the model, and are robust across the range of physical parameter values realistic for natural environments (Fig. 5). In particular, varying the key metabolic parameters of our model by as much as an order of magnitude above and below our default values has a relatively small impact on our results (Fig. 5a). In addition, there is no *a priori* reason to expect that the basic ecophysiological phenomenon underlying our results — effective photoferrotrophic consumption of nutrients at low light levels — has changed dramatically over time through the course of microbial evolution (see below). Our conclusions are thus most strongly dependent on the physical parameters used in the model. For example, relative oxygenic phototrophy can be increased appreciably relative to our default simulation by decreasing light penetration (e.g., increasing λ; Fig. 5b), increasing rates of mixing due to eddy diffusion ($K_v$; Fig. 5c), or decreasing upwelling rates ($w$; Fig. 5e). However, in all of these cases the values required to markedly impact our default oxygenic productivity are very extreme relative to natural surface ocean settings[44-49]. In addition, these changes could be easily offset by reasonable increases[23] in the distribution coefficient for P scavenging ($K_d$; Fig. 5c), which will be regulated at any given period of Earth's history by seawater concentrations of dissolved Si, Ca, and Mg. The basic framework implied by our water column ecology model is thus very robust to reasonable variability in both metabolic and physical parameters.

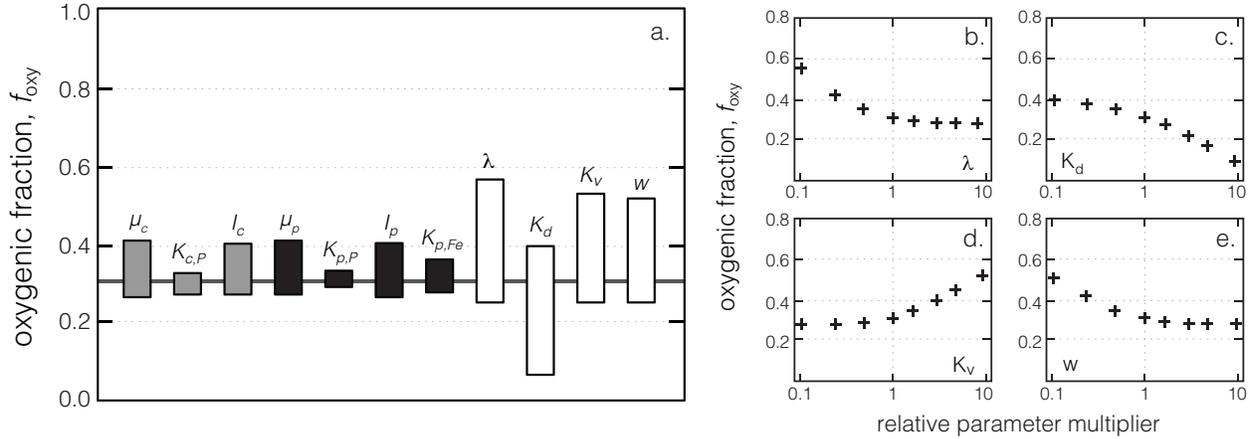

**Figure 5. Sensitivity analysis of the 1-D competitive photosynthesis model.** Shown in (**a**) is the response of our principle diagnostic parameter, the fraction of overall photosynthesis performed by oxygenic phototrophs ($f_{oxy}$), to varying each metabolic and physical parameter of the model by an order of magnitude above and below the value in the default parameter set (shown by the black horizontal line and denoted by the open circle in Figure 4). Major metabolic parameters are maximum growth rates ($\mu_i$) and half-saturation constants for light ($I_i$) and nutrients ($K_{i,P}$) for oxygenic phototrophs and photoferrotrophs (subscripts $c$ and $p$, respectively). Physical parameters include light attenuation constant ($\lambda$), distribution coefficient for scavenging/coprecipitation of phosphorus on iron oxides ($K_d$), eddy diffusivity ($K_v$), and upwelling rate (w). Shown in (**b**)-(**e**) are sensitivity ensembles for physical parameters, with relative oxygenic phototrophy plotted as a function of relative forcing of each parameter.

**Global ocean-sediment biogeochemical model: CANOPS-KB.** In order to explore the potential large-scale impacts of competitive ecophysiology on Earth's oxygen cycle, we employ an Earth system model (ESM) of the coupled C-N-P-O$_2$-S cycles. Our model, CANOPS-KB (see Methods), has a general and robust circulation scheme that is capable of producing well-resolved distributions of $^{14}$C and temperature and has been extensively tested and validated against observations from the modern and ancient Earth[50,51]. The biogeochemical scheme is based on a primary limiting nutrient, phosphate (PO$_4^{3-}$), that ultimately controls biological productivity in the surface ocean, and includes biological productivity in the low- and high-latitude surface ocean, a series of respiratory pathways in the ocean interior (aerobic respiration, denitrification, and sulphate reduction), secondary redox reactions (nitrification and aerobic sulphide oxidation), and the deposition, processing, and burial of organic matter and phosphorus in marine sediments. We also include redox-dependent scavenging (and subsequent burial) of phosphorus by Fe oxides building on previous work[27] (see Methods), and ecological competition between photoferrotrophs and oxygenic photosynthesizers based on a simplified version of our 1-D water column model (see Methods).

By specifying atmospheric $O_2$, external inputs of P, and the ratio between dissolved Fe(II) and $PO_4^{3-}$ in the deep ocean ($[Fe/P]_d$), we can use the model to compute the global burial rate of organic carbon originating from oxygenic phototrophs — and thus estimate the overall production of $O_2$ by the biosphere on geologic timescales. We can then combine this with an estimation of $O_2$ sinks due to inputs of volcanic/metamorphic reductants and oxidative weathering (the latter of which scales with atmospheric $O_2$; see Methods) to calculate the net $O_2$ balance at Earth's surface ($\Phi_{ox}$). When biospheric $O_2$ production fluxes (i.e., the burial of organic carbon produced by oxygenic phototrophs) and consumption fluxes due to weathering and volcanic/metamorphic inputs are equal ($\Phi_{ox} = 0$), and the global $O_2$ cycle is in equilibrium (open/closed circles in Fig. 6a). When $\Phi_{ox} > 0$ the biosphere is releasing $O_2$ at rates that exceed available sink fluxes, and atmospheric $O_2$ should rise. Similarly, when $\Phi_{ox} < 0$ consumption rates are greater than rates of production by the oxygenic biosphere, and atmospheric $O_2$ should fall.

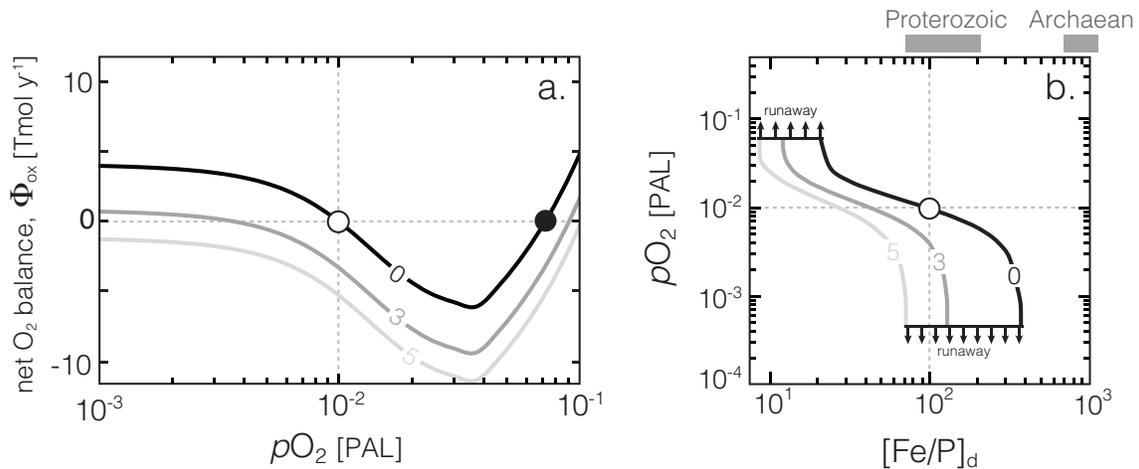

**Figure 6. Representative results from our global ocean-sediment biogeochemical model (CANOPS-KB).** (**a**) Each curve shows net oxygen fluxes ($\Phi_{ox}$) as a function of atmospheric $pO_2$. Intersection points at $\Phi_{ox} = 0$ indicate regions of parameter space in which the $O_2$ cycle is at equilibrium and dominated by negative feedback (negative slope) and positive feedback (positive slope). Circles plotted along the black curve thus show examples of stable (open) and unstable (closed) equilibrium points for model runs at a nominal deep ocean Fe/P ratio of 100. Each curve is labeled by an assumed external reductant flux (i.e., volcanic/metamorphic reductant input minus $O_2$ accumulation due to hydrogen escape and/or imbalances in global S and Fe cycling, $\Phi_{red}$) in Tmol $O_2$ equivalents $y^{-1}$. (**b**) Each curve plots stable equilibrium $pO_2$ values as a function of deep ocean Fe/P ratio ($[Fe/P]_d$) under the same conditions as shown in (**a**). The open circle shows our nominal model ($[Fe/P]_d = 100$ and no external reductant input). Arrows above and below the curves denote regions where our model predicts runaway oxygenation or deoxygenation, respectively. Also shown in (**b**) are approximate estimates of $[Fe/P]_d$ for the Archaean and Proterozoic (shaded grey bars; see text).

We can also use this analysis to evaluate the stability of a given equilibrium point[52]. For example, if the slope at equilibrium is negative (closed circle in Fig. 6a), there is negative feedback and the equilibrium point will be relatively stable to external perturbation. Conversely, if the slope at equilibrium is positive (open circle in Fig. 6a), there is positive feedback and the equilibrium point should be unstable. The gradient of the slope indicates the overall stability at equilibrium. It is important to bear in mind that our analysis specifies [Fe/P]$_d$ as a boundary condition, while in reality we should expect this parameter to scale in some fashion with ocean-atmosphere $O_2$ abundance. This scaling is currently unknown and awaits more comprehensive knowledge of the coupled Fe-P-C-S-$O_2$ cycles in pervasively anoxic oceans. Nevertheless, we consider this approach useful for exploring the role of anoxygenic photosynthesis in the stability of Earth's oxygen cycle.

Our stability analysis of the Earth surface oxygen cycle including competitive phototrophy yields two key insights. First, the location of stable equilibria depends strongly on [Fe/P]$_d$ – for example, for our benchmark model (with [Fe/P]$_d$ = 100 and no external reductant flux) there is a stable equilibrium point at an atmospheric $pO_2$ value of ~2 ×10$^{-3}$ atm, or ~1% PAL (open circle in Fig. 6a). However, the $pO_2$ value that yields a stable equilibrium scales directly with [Fe/P]$_d$ (Fig. 6b), such that higher [Fe/P]$_d$ yields lower $pO_2$ and vice versa. This effect can be further modulated by changes in external volcanic/metamorphic reductant fluxes. For example, the $pO_2$ corresponding to stable equilibrium drops to ~0.3-0.4% PAL in our benchmark model if we impose an external reductant flux of 3 Tmol $O_2$ equivalents y$^{-1}$ (Fig. 6a), on the low end of estimated volcanic/metamorphic reductant fluxes to modern Earth surface environments[53].

Second, our analysis predicts a threshold [Fe/P]$_d$ value below which the system will undergo runaway oxygenation (Fig. 6b). In our benchmark model this occurs at [Fe/P]$_d$ of ~20, below which enhanced nutrient fluxes allow oxygenic phototrophs to oxygenate the photic zone, suppressing photoferrotrophic activity and phosphorus scavenging by iron oxides, promoting eutrophication of the ocean interior, and further enhancing nutrient fluxes to oxygenic phototrophs. Such a runaway oxygenation could occur as a result of increasing oceanic P or through a sufficient long-term drop in reductant flux. The stable $pO_2$ value also becomes highly sensitive to [Fe/P]$_d$ when [Fe/P]$_d$ > 200 (Fig. 6b), with relatively small changes in [Fe/P]$_d$ causing large changes in the $pO_2$ at which stable equilibrium is possible. By contrast, in our nominal model [Fe/P]$_d$ values approaching ~400 strongly suppress the activity of oxygenic phototrophs. Although our model does not currently include the photochemical reaction network linking molecular oxygen ($O_2$), ozone ($O_3$), and methane ($CH_4$) in the atmosphere[54,55], it is anticipated that such a strong suppression of the activity of the oxygenated biosphere would tip the system into a runaway deoxygenation and result in a stable, reducing atmosphere (Fig. 6b).

**Discussion**

We argue that strain KB01 represents a strong physiological analogue to the ancestral photoferrotrophs that would have populated ferruginous Precambrian oceans. The ecological prominence of the Chlorobiales in many modern analogue environments implies that they are well adapted to growth and proliferation under conditions relevant to Precambrian marine environments[56]. At the same time, the deep ancestry of the Chlorobiales implies that the genetic underpinning of this physiology has been conserved across billions of years—as is the case for N-fixation[56]. While it is challenging to connect relative antiquity implied from sequence analyses (Fig. 3) to absolute age dates that can be tethered to the geologic record, biomarkers diagnostic of ancestral Chlorobiales have been recovered from 1.6 Ga shales[57], while molecular clocks imply emergence of stem group Chlorobi as early as the Mesoarchaean[58]. This implies that our results are likely applicable to the Mesoarchaean or earlier and are almost certainly relevant to periods of Earth's history subsequent to the early Paleoproterozoic.

In addition to nutrient metabolism[56] (see above), both pigment biosynthesis pathways[10] and accessory components of low-light photosystems[59] are strongly conserved within photosynthetic lineages. For example, the Fenna-Mathews-Olson (FMO) complex, which mediates energy transfer from chlorosomes to the photosynthetic reaction center and is unique to the Chlorobiales[60], is phylogenetically and functionally conserved[59,61], suggesting it was likely present in the last common ancestor of crown group Chlorobiales. The relative antiquity of the FMO complex and its role in bridging chlorosomes to the reaction center, which confers exceptional light harnessing capacity to the Chlorobiales, thus implies that the last common ancestor of crown group Chlorobiales was very well adapted to growth under low-light conditions and that this capacity emerged in stem group photosynthetic Chlorobi as early as the Mesoarchaean[58], or possibly earlier. In addition, photoferrotrophy by ancestral members of other photosynthetic lineages could also have played a similar ecological role and indeed high affinity $PO_4^{3-}$-uptake systems are broadly distributed across the bacteria[62]. Taken as a whole, sequence information, the ecology of extant Chlorobiales, and the geologic record all support the antiquity of the nutrient and light metabolism displayed by strain KB01.

The dynamics of our large-scale biogeochemical model depend strongly on the abundances of dissolved Fe and P in the ocean interior during Archaean and Proterozoic time, which are not well-constrained at present. Nevertheless, we can obtain a basic sense of the links between our model and evolving ocean chemistry from constraints derived from laboratory experiments, thermodynamic calculations, and observations from the rock record. Recent estimates[63,64] based largely on the hypothesis of a metastable green rust phase as a primary Fe mineral in Precambrian marine systems place Archaean dissolved Fe on the order of ~100 µmol kg$^{-1}$, with Proterozoic dissolved Fe roughly an order of magnitude lower[64]. These estimates are broadly consistent with the observation of siderite in Archaean carbonate platforms[65] and the long-term thermodynamic stability of Fe-silicate phases in Archaean seawater[66], though values for the Proterozoic could have been considerably higher than this if the formation of Fe-bearing carbonate and silicate phases was kinetically inhibited[16].

If we combine these values for dissolved Fe(II) with estimates of deep ocean P concentration based on the P/Fe ratios observed in iron-rich chemical sediments deposited during the Archaean and Proterozoic[23], we obtain [Fe/P]$_d$ values for the Archaean of roughly 700 – 2,000 and Proterozoic values around an order of magnitude below this. Interpreted in the context of our model, these values would imply an essentially anoxic Archaean atmosphere despite the presence of oxygenic phototrophs in the surface oceans (Fig. 6b). We find that this is true even with no additional reductant flux from volcanic/metamorphic gases — an important observation given that proposed mechanisms for the first widespread oxygenation of Earth surface environments at ~2.3 Ga (the Great Oxidation Event, or GOE) overwhelmingly gravitate toward the role of geophysical $O_2$ sinks[67-72]. Our results thus clearly demonstrate the importance of microbial physiology and deep ocean chemistry as controls on atmospheric $O_2$ and highlight the previously unrecognized role of deep ocean Fe:P ratios as a potential throttle on the timing and intensity of the GOE.

Furthermore, these estimates would yield a largely stable low-oxygen atmosphere during the Proterozoic according to our model (Fig. 6b). Uncertainties in both reductant fluxes from the solid Earth and deep ocean chemistry prevent us from definitively assigning any particular simulation to specific intervals of Proterozoic time, and it remains possible that ocean-atmosphere $O_2$ levels were variable throughout this interval as a function of time-dependent changes in rates of volcanism, ocean $[Fe/P]_d$ values, and other factors. However, the $pO_2$ stability range in our model under roughly Proterozoic conditions is consistent with observations from the rock record, including the Fe/Mn chemistry of ancient weathering profiles[73-75] and the stable isotope compositions of iron-rich chemical sediments[76], shales[77], and marine evaporite minerals[74,78]. Other estimates placing Proterozoic $pO_2$ at higher values[79-81] are also broadly consistent with our model[82,83], which may provide a mechanism that would allow a range of stable atmospheric oxygen levels encompassing existing geochemical estimates as well as transitions between these levels on relatively short time scales. In any case, taken together with the apparent predominance of Fe(II)-rich conditions in the ocean interior throughout the Precambrian[13,22,84], our results implicate competition between different modes of photosynthesis as important in regulating the oxygen cycle during large intervals of Earth's history.

In summary, our coupled experimental-modeling results suggest an important role for competitive ecophysiology, modulated by changes in deep ocean chemistry, in controlling Earth's surface $O_2$ cycle. More specifically, we provide a simple biological mechanism for preventing atmospheric oxygenation in the face of sustained photosynthetic $O_2$ production during the Archaean Eon and for explaining low atmospheric $pO_2$ during the Proterozoic Eon. We also find a potential positive feedback within the coupled C-P-O-Fe cycles that can lead to runaway planetary oxygenation, as rising atmospheric $pO_2$ sweeps the deep ocean of the ferrous iron substrate for photoferrotrophy, stimulates expansion of the oxygenic component of the photosynthetic biosphere, and causes atmospheric $pO_2$ to rise further. Future advances in quantitatively constraining ocean-atmosphere chemistry will help to pinpoint the external forcings required to initiate this runaway oxygenation, and definitively establish the conditions under which a similar runaway deoxygenation might occur. Nevertheless, our results demonstrate the fundamental importance of photosynthetic ecology and deep ocean Fe:P ratios for Earth system evolution.

## Methods

**Metabolic potential of *C. phaeoferrooxidans* strain KB01**

*Chlorobium phaeoferrooxidans* strain KB01 was grown in medium adapted from Ref[85], where potassium phosphate was replaced with potassium chloride. 0.6 mM Si was added to the medium and the ferrous iron concentration was 1.4 mM. The growth medium was inoculated with *C. phaeoferrooxidans* strain KB01 that had been previously grown on medium containing 4 mM phosphate. To limit P carryover through the inoculum, it was first centrifuged (5 minutes at 16,100 rcf) and the supernatant discarded prior to introduction to the P-free medium. Once inoculated, radiolabeled phosphate ($^{32}$P) was added to each bottle so that the final phosphate concentration was approximately 1.5 µM with radiation levels of 2 mCi.

To assess growth and P uptake kinetics, time points were taken every 3-5 hours for 88 hours. Subsamples were collected at each time point for iron speciation, pigment analysis, total radioactivity, dissolved radioactivity, radioactivity incorporated into the biomass, and radioactivity bound to ferric iron particles or cellular biomass. The iron speciation measurements were conducted using the ferrozine assay[86], while the pigments were extracted using a 7:2 methanol:acetone mixture[87]. The sample for dissolved radioactivity was measured after filtration to 0.2 µm by scintillation counting of 1 mL of filtrate, while the total radioactivity was measured without filtration. The $^{32}$P incorporated into biomass was determined by centrifuging 0.5 mL of sample for 5 minutes at 16,100 rcf, discarding the supernatant, and then treating with dithionite for 15 minutes to remove particulate iron and associated P. After treating with dithionite, the samples were centrifuged again for 5 minutes at 16,100 rcf, supernatant was discarded and 1 mL of iron-free media (no phosphate and containing 0.6 mM Si) was added to the sample, which was then counted. The $^{32}$P sorbed to particles in solution was counted by treating 0.5 mL of sample with 1 mL of dithionite for 15 minutes. 0.5 mL of iron-free media (no phosphate and 0.6 mM Si) was then added to the sample and 1 mL of this was filtered to remove biomass and counted.

To determine metabolic potential for $PO_4^{3-}$ metabolism in photoferrotrophic Chlorobiales, we searched the previously annotated genome of strain KB01 for genes known and implicated in microbial responses to P-starvation. The protein identities were confirmed with a BLAST[88] search against the NCBI non-redundant database. Recovered genes were placed into their broader genomic context by searching for the host contig and recovering adjacent genes. The evolutionary history of $PO_4^{3-}$ metabolism in the Chlorobiales was assessed through comparative phylogenetic analyses of the proteins PstA, PstB, and PstC, in a concatenated matrix, as well as PhoU, versus the 16S rRNA gene. Sequences were aligned using ClustalX2.1[89], and phylogenetic trees constructed with multiple methods (Neighbor Joining, FastTree, and verified with Maximum Likelihood and Maximum Parsimony). Tree nodes are labeled with both bootstrap (Neighbor Joining) and FastTree support values.

**1-D water column model of competitive photosynthesis**

Our 1-D water column model solves an advection-diffusion-reaction equation of the form:

$$\frac{\partial C}{\partial t} = w\frac{\partial C}{\partial z} - K_v \frac{\partial^2 C}{\partial z^2} + \sum R_i \quad , \quad (1)$$

where $C$ represents dissolved concentrations (e.g., $Fe^{2+}$, $PO_4^{3-}$), $z$ is depth, $w$ is advection velocity (e.g., upwelling rate), $K_v$ represents a diapycnal eddy diffusivity, and $R_i$ denotes a series of kinetic reaction terms that control a given dissolved species in the water column. The equation is solved via finite-difference techniques using the R package ReacTran[90].

Our model specifies a simple form of competition between photoferrotrophs and oxygenic photosynthesizers based qualitatively on our experimental work with *C. phaeoferrooxidans* str. KB01. Photoferrotrophic growth is parameterized as a function of ambient light availability, dissolved ferrous iron levels, and dissolved nutrients (e.g., $PO_4^{3-}$), while cyanobacterial growth is constrained by both light and dissolved nutrients:

$$J_{photoferro} = \mu_p \cdot \frac{I_z}{I_p + I_z} \cdot \frac{[Fe^{2+}]}{K_{p,Fe} + [Fe^{2+}]} \cdot \frac{[PO_4^{3-}]}{K_{p,P} + [PO_4^{3-}]} \quad , \quad (2)$$

$$J_{cyano} = \mu_c \cdot \frac{I_z}{I_c + I_z} \cdot \frac{[PO_4^{3-}]}{K_{c,P} + [PO_4^{3-}]} \quad , \quad (3)$$

where $\mu_i$ and $K_i$ terms represent maximum growth rates and nutrient half-saturation constants for each photosynthetic metabolism, $I_i$ terms represent light half-saturation constants for each photosynthesizer, and $I_z$ represents light availability at a given depth in the water column. Light availability is assumed to vary with depth according to a simple exponential decay:

$$I_z = I_0 e^{-\lambda z} \quad , \quad (4)$$

where $I_0$ represents incident light at the surface, $z$ represents depth, and $\lambda$ represents an attenuation length-scale.

Inorganic oxidation of dissolved $Fe^{2+}$ with $O_2$ proceeds according to:

$$J_{ox} = k_{ox}[Fe^{2+}][O_2][OH^-]^2 \quad , \quad (5)$$

where the kinetic rate constant is a function of temperature ($T$) and ionic strength ($s_i$) following Ref. 43:

$$\log k_{ox} = \left[21.56 - \frac{1545}{T}\right] - 3.29 s_i^{0.5} + 1.52 s_i \quad , \quad (6)$$

Finally, scavenging and co-precipitation of inorganic $PO_4^{3-}$ onto/into Fe-oxide mineral phases occurs according to a simple distribution coefficient ($K_d^{FeP}$). All default model parameters are given in Supplementary Table 2.

**Global ocean-sediment biogeochemical model: CANOPS-KB**

We employ a 1-D (vertically resolved) transport-reaction model of marine biogeochemistry to explore the global impact of ecophysiological constraints on the coupled Earth surface C-N-P-$O_2$-S cycles. Our model, CANOPS-KB (Supplementary Figure 1), is an improved version of the original CANOPS model developed by Ozaki and Tajika[91], and subsequently modified by Reinhard et al.[92]. The details of base model configuration, parameterization and validation are described elsewhere[91,92], but here we outline the basic model structure and the modifications developed for this study.

The CANOPS framework has a general and robust circulation scheme that is capable of producing well-resolved distributions of circulation tracers (e.g., $^{14}C$ and temperature). The diffusion-advection model of the global ocean is coupled with a biogeochemical model and a parameterized sediment model. The biogeochemical scheme is based on a primary limiting nutrient, phosphate ($PO_4^{3-}$), that ultimately controls biological productivity in the surface ocean, and includes biological productivity in the low- and high-latitude surface ocean, a series of respiratory pathways (aerobic respiration, denitrification, and sulphate reduction), secondary redox reactions

(nitrification and aerobic sulphide oxidation), and deposition and burial of organic matter and phosphorus in marine sediments. We also include redox-dependent scavenging (and subsequent burial) of phosphorus by Fe oxides[92]. The principal biogeochemical processes, formulations, and constants used in the model are summarized in Supplementary Tables 3-5, and the major process modifications are summarized here.

Ecological competition between photoferrotrophs and oxygenic photosynthesizers in our large-scale model is informed by our metabolic experiments with *C. phaeoferrooxidans* str. KB01 and the results of our 1-D water column model. In the CANOPS-KB model, the total new production ($J_{ex}^{total}$; in mol C m$^{-2}$ y$^{-1}$) is assumed to be intrinsically phosphorus limited;

$$J_{ex}^{total} = \alpha \cdot h_m \cdot \varepsilon \cdot [PO_4^{3-}] \cdot \frac{[PO_4^{3-}]}{K_m^P + [PO_4^{3-}]} , \quad (7)$$

where $\alpha$ represents C/P stoichiometry of producers (=106), $h_m$ is the surface layer depth, and $\varepsilon$ is the efficiency factor for phosphorus uptake, and $K_m^P$ denotes the Monod half-saturation parameter for phosphate-limited growth. In this study we assume that photosynthetic carbon fixation adheres to canonical Redfield stoichiometry (C:N:P = 106:16:1) and that the nitrogen required to sustain productivity is compensated by the activity of diazotrophs. The phosphate availability in the surface layer is explicitly calculated based on the riverine input, water transport via advection and diffusion, and biological production. New production of photoferrotrophic biomass is determined by the relative availability of both Fe(II) and PO$_4^{3-}$. Under conditions of Fe-limited growth within the photoferrotrophic community, any P remaining after complete consumption of Fe(II) can then be used by oxygenic photosynthesizers (e.g., cyanobacteria). Under P-limited photoferrotrophic growth, all bioavailable phosphorus is consumed by photoferrotrophs and the contribution of oxygenic photosynthesizers to new production is curtailed. We can express the rates of new production due to photoferrotrophy and oxygenic photosynthesis in terms of the fluxes of particulate carbon exported from the euphotic zone ($J_{ex}$; in mol C m$^{-2}$ y$^{-1}$) as follows:

$$J_{ex}^{pfe} = \min\left[\frac{J_{Fe}^{up}}{r_{FeC}}, J_{ex}^{total}\right] , \quad (8)$$

$$J_{ex}^{cyano} = \begin{cases} J_{ex}^{total} - J_{ex}^{pfe} & : J_{ex}^{pfe} < J_{ex}^{total} \\ 0 & : J_{ex}^{pfe} = J_{ex}^{total} \end{cases} , \quad (9)$$

where *pfe* and *cyano* represent contributions due to photoferrotrophs and oxygenic photosynthesizers (e.g., cyanobacteria), respectively, $r_{FeC}$ represents the stoichiometric ratio between Fe and C attendant to photoferrotrophy, and $J_{Fe}^{up}$ is the upward flux of Fe(II) to the photic zone via advection and turbulent diffusion given by:

$$J_{Fe}^{up} = [Fe/P]_d \cdot J_P^{up} \quad , \quad (10)$$

where the upward $PO_4^{3-}$ flux, $J_P^{up}$, is given by

$$J_P^{up} = A_{j=1} \cdot w \cdot [PO_4^{3-}]_{j=1} + A_{j=1} \cdot K_v \cdot \left. \frac{\partial [PO_4^{3-}]}{\partial z} \right|_{z=h_m} \quad , \quad (11)$$

for the low-mid latitude surface region, where $w$ (m y$^{-1}$), $K_v$ (m$^2$ y$^{-1}$), and $A_{j=1}$ are upwelling velocity (in m y$^{-1}$), vertical turbulent diffusivity (5000 m$^2$ y$^{-1}$), and areal fraction of the $j=1$ layer to the surface area of low-mid latitude region, respectively. $h_m$ (m) is the depth of the mixed layer.

The scavenging and co-precipitation of inorganic $PO_4^{3-}$ onto/within Fe-oxide mineral phases (in terms of mol P m$^{-2}$ y$^{-1}$) is formulated according to a simple distribution coefficient ($K_d^{FeP}$):

$$J_{scav} = \begin{cases} \gamma \cdot K_d^{FeP} \cdot [PO_4^{3-}]_l \cdot J_{Fe}^{up} & \text{when } [O_2]_{j=1} < 1\,\mu M \\ 0 & \text{when } [O_2]_{j=1} \geq 1\,\mu M \end{cases} \quad , \quad (12)$$

where γ denotes the ultimate burial efficiency of scavenged P. The activity of photoferrotrophs (and scavenging) is specified to cease when the layer just below the surface layer ($j$=1) becomes oxygenated.

**Rates of oxidative weathering of organic matter**

In order to explore the potential consequences of photoferrotrophy for the stability of Earth's global oxygen cycle, we employ a simple parameterization of the rate of oxidative weathering of organic matter in the crust ($J_w^{org}$) as a function of atmospheric $pO_2$ according to:

$$\frac{J_w^{org}}{J_{w,0}^{org}} = f_w \cdot \exp\left[-\sigma_{org}^1 \cdot \exp[-\sigma_{org}^2 \cdot PAL]\right] \quad , \quad (13)$$

where $\sigma_{org}^i$ terms refer to fitting constants (here with values of 4 and 150, respectively), PAL refers to atmospheric $pO_2$ relative to the present atmospheric level, and $J_{w,0}^{org}$ refers to the modern globally integrated organic carbon oxidation rate. This relationship is based on an approximate fit to results from a 1-D reaction-transport model of organic carbon weathering in the crust[93,94]. We introduce a free parameter in Eq. (13), $f_w$, which expresses the effect of continental erosion rate on the global rate of oxidative weathering ($f_w = 1$ for the benchmark simulations presented in the Main Text). We also assume that the riverine P input flux and sediment accumulation rate at the seafloor (*SR*) scale in a linear fashion with $f_w$. We employ this method for simplicity but note that the location and overall stability of equilibrium $pO_2$ values will respond to some extent on the parameters employed in the reaction-transport model (Supplementary Figure 2). However, the basic components of our argument remain unchanged.

**Evaluating the exogenic oxygen budget**

We can define a global redox budget (in $O_2$ equivalents) for the combined ocean-atmosphere system as:

$$, \qquad (S1)$$

where $\Phi_{ox}$ denotes the net redox flux balance within the ocean-atmosphere system, $\Phi_{org}$ denotes oxygen production through the burial of oxygenic photosynthetic biomass in marine sediments, $\Phi_{weath}$ denotes the consumption of $O_2$ from the atmosphere during the oxidative weathering of organic carbon in the crust, and $\Phi_{red}$ denotes the net input flux of reducing power via the combined effects of volcanic/metamorphic gas input[53,95], escape of hydrogen to space, and any redox imbalance within the S and Fe budgets.

In our default analysis of the $O_2$ budget we do not explicitly evaluate $O_2$ fluxes associated with the pyrite sulphur subcycle or hydrogen escape from the upper atmosphere[67,96,97]. In the case of the pyrite sulphur subcycle, the isotope record of sedimentary sulphur-bearing minerals for most of Precambrian time is conventionally considered to suggest that effectively all sulphur entering the oceans was removed as a constituent of sedimentary pyrite. In this case, $O_2$ fluxes within the pyrite sulphur subcycle should be largely balanced on the longest timescales, though we note that the oxidative weathering of pyrite potential provides an additional stabilizing mechanism for Earth's $O_2$ budget[98] and that the volcanic outgassing of $SO_2$ and subsequent burial of pyrite may also make a non-trivial contribution to overall redox balance[95,99]. The flux associated with H escape from the atmosphere is very small at present, but may have been significantly larger during periods of elevated atmospheric $CH_4$ levels, in particular during the Archaean Eon. We can combine this flux with the input flux of reductants from the solid Earth, and consider this combined exogenic $O_2$ flux as part of the reductant flux boundary condition ($\Phi_{red}$) discussed in the Main Text. However, we stress that the conceptual framework we present for Earth's oxygen cycle is a natural and robust consequence of the inclusion of photoferrotrophy, and that this would not be altered by considering a more inclusive $O_2$ budget. Nevertheless, our results highlight the importance of time-dependent stability analysis of the exogenic $O_2$ cycle as an important topic for future work.

**Data availability.**

All analytical data not given in the Supplementary Information are available on request.

**Code availability.**

Code for the 1-D water column competitive photosynthesis model is written in R and is available on GitHub (https://github.com/ChrisReinhard/photoferrotrophy.R). Code for the CANOPS-KB model is written in Fortran and is available from K.O. or C.T.R. upon reasonable request.

**Acknowledgements**

K.O. acknowledges support from the NASA Postdoctoral Program at the NASA Astrobiology Institute, administered by Universities Space Research Association under contact with NASA. K.O. also acknowledges JSPS KAKENHI Grant Number JP25870185. C.T.R. acknowledges support from the NASA Astrobiology Institute, the National Science Foundation, and the Alfred P. Sloan Foundation. This work was also supported by NSERC Discovery Grant 0487 to S.A.C. Steven Hallam, Aria Hahn, and Martin Hirst helped generate genomic data for strain KB01.

**Author contributions**

K.O., S.A.C., and C.T.R. designed the research. K.O., C.T.R., and S.A.C. designed and implemented the biogeochemical models. K.J.T. performed physiological and microbiological analysis of *C. phaeoferrooxidans* KB01. R.L.S., K.J.T., and S.A.C. performed genomic and phylogenetic analyses. All authors contributed to interpretation of the combined results and the writing of the manuscript.

**Competing interests:** The authors declare no competing interests.